\documentclass[aps,preprint,showpacs]{revtex4}

\usepackage{graphicx}

\newcommand{\comment}[1]{}

\begin{document}

\title{Spin and charge optical conductivities in spin-orbit coupled systems}

\author{Jes\'us A. Maytorena{\email{jesusm@ccmc.unam.mx}},
        Catalina L\'opez-Bastidas, and Francisco Mireles}

\affiliation{Centro 
de Ciencias de la Materia Condensada, Universidad Nacional 
Aut\'onoma de M\'exico, Apdo. Postal 2681, 22800 Ensenada, 
Baja California, M\'exico}

\begin{abstract}
\vspace{0.1 in}
We study the frequency dependent spin- and charge- conductivity tensors of a
two-dimensional electron gas (2DEG) with Rashba and Dresselhaus spin-orbit
interaction. 
We show that the angular anisotropy of the spin-splitting energy induced by
the interplay between the Rashba and Dresselhaus couplings gives rise to a 
characteristic spectral behavior of the spin and charge response which is 
significantly different from that of pure Rashba or Dresselhaus case. Such 
new spectral structures open the possibility for control of the optical 
response by applying an external bias and/or by adjusting the light
frequency. In addition, it is shown that the relative strength of the 
spin-orbit coupling parameters can be obtained through optical probing.  

\end{abstract}

\pacs{73.63.Kv, 72.25.-b, 73.21.La, 72.25.Dc}

\date{\today}
\maketitle
Spin-orbit interaction (SOI) in systems lacking inversion symmetry is a 
phenomenon with great potential in the development of spintronic-based
devices. Since the celebrated proposal\cite{DattaDas} of a spin-FET relying 
in the tunability of the Rashba SOI strength through electrical gating, 
there has been a remarkable attention in the search for new ways of 
manipulating electron spins without employing ferromagnetic materials 
and/or external magnetic fields. For instance, a spin Hall effect in 
which a transverse spin current is driven by a dc electric field 
(without a net charge current) was predicted to arise in low-dimensional 
systems with a substantial SOI. 
\cite{Sinova,Murakami,Sinova-cond-mat,Schliemann-cond-mat}
Spin (Hall) accumulation has been observed through optical measurements 
\cite{Wunderlich, Kato,Sih} and very recently the first observation, 
purely electrical, of the spin Hall effect in a metalic conductor was 
reported \cite{Valenzuela}.  

The spin splitting $\Delta$, caused by SOI in electron systems, opens the 
possibility of resonant effects as a response to alternating electric fields
due to transitions between the  spin-split states.\cite{Entin,Sipe,Rashba}
For instance, the absorption of linearly polarized infrared radiation has been 
recognized as a mechanism to inject pure spin currents in quantum wells by 
inter-spin-split-subband transitions.\cite{Sipe}
An ac spin current generation by the time variation of the Rashba coupling
through time modulated gate voltages has also been proposed. 
\cite{Malshukov,Governale} In a recent study it has been suggested that an 
intense ac probing field  can be used to control the spin-Hall current in 
2DEGs with Rashba SOI.\cite{Wang} 
It has also been shown that in the THz range of frequencies, the cancellation 
of the instrinsic spin Hall effect due to impurity scattering \cite{Khaestkii} 
is no longer perfect, and in principle the effect could be
present.\cite{Finkelstein} Moreover, for finite frequencies at the window of 
$\Delta > \hbar \omega >\hbar \tau^{-1}$, in which $\tau^{-1}$ is the impurity 
scattering rate, the spin Hall conductivity converges to its universal 
intrinsic value of $e/8\pi$.\cite{Mishchenko, Chalaev-Loss,Grimaldi}
Another resonant phenomenon is the electric-dipole-induced spin resonance
in which the electron spins may be manipulated via SO coupling by means of time-dependent
electric fields (rather than resonant magnetic fields) as is shown to occur 
in both clean \cite{Efros}
and disordered \cite{Loss-EDSR} systems.

Other recent studies have  also emphasized the importance of the dynamical
regime, \cite{Zhang,Halperin} investigating  several relevant physical aspects, 
such as the relation between the spin Hall conductivity and 
the spin susceptibility\cite{Erlingsson,Finkelstein} or the
dielectric function,\cite{Rashba} the presence of electron-electron 
interactions and plasmon modes\cite{Entin,Xu03,XFWang,Pletyukhov}, the study of the 
optical absorption spectrum \cite{Xu05} and the renormalization of the Rashba 
parameter and the spin-splitting energy.\cite{Finkelstein}
 More recently, effects of strains\cite{Cheche} 
and of the electron-phonon interaction\cite{Grimaldi} on the spin Hall 
currents have been also explored.   All these studies considered only 
the Rashba type of SOI. However, both experimental\cite{Ganichev,Miller} 
and theoretical work  \cite{Halperin, Loss, Egues, Erlingsson,Sinitsyn,SQShen} 
have pointed out the importance of  the (linear) Dresselhaus type of SOI 
contribution and its interplay with the Rashba coupling.

In this work, we study the charge- and spin-current conductivity tensors of a
2DEG with Rashba and Dresselhaus SOI as a linear response to a frequency 
dependent (spatially homogeneous)  weak electric field. We find that the 
angular anisotropy of the energy spin-splitting introduced by the interplay 
between both coupling strengths yields a finite-frequency response with 
spectral features that are  significatively different from that of a pure 
Rashba (Dresselhaus) coupling  case. As a consequence, an optically modulable 
spin  and charge current response is then achievable in such systems. It is 
noticed as well that the inter-spin-split subband excitations via photon 
absorption are now possible in a wider frequency range, depending explicitly 
upon the Fermi wave vector and on the SOI
parameters. Futhermore, it is shown that such effect may be used to extract 
the ratio between the SOI coupling parameters via optical and/or transport 
experiments.

We consider a 2D free electron system lying at $z=0$ plane, with a
Hamiltonian given by $H=\frac{\hbar^2k^2}{2m^*} + H_{so}$, 
where the spin-orbit interaction is
\begin{equation} 
\label{h_so}
H_{so}=\alpha(k_x\sigma_y-k_y\sigma_x)+\beta(k_x\sigma_x-k_y\sigma_y)\ \ ,
\end{equation}

The first term corresponds to the Rashba SO coupling which originates from any
source of structural inversion asymmetry of the confining potential. 
The second term is the linear Dresselhaus coupling which results from 
bulk-induced inversion asymmetry. The spectral properties of this Hamiltonian 
are well known.\cite{Halperin,Loss,Egues} The eigenstates 
$|{\bf k}\lambda\rangle$ for the in-plane motion are specified by the 
wave vector ${\bf k}=(k_x,k_y)=k(\cos\theta,\sin\theta)$ and  chirality 
$\lambda=\pm 1$ of the spin branches.  The double sign corresponds to 
the upper (+) and lower ($-$) parts of the energy spectrum given by 
$\varepsilon_{\lambda}(k,\theta)=\hbar^2k^2/2m^{*}+\lambda\,k \Delta(\theta)$, 
where $\Delta(\theta)=\sqrt{\alpha^2+\beta^2-2\alpha\beta \sin2\theta}$ 
describes the angular anisotropy of the spin splitting. At zero temperature, 
the two spin-split subbands are filled up to the same (positive) Fermi energy 
level $\varepsilon_F$ but with different Fermi wave vectors  
$q_{\lambda}(\theta)=\sqrt{2m^*\varepsilon_F/\hbar^2+k_{so}^2(\theta)} 
-\lambda k_{so}(\theta)$, determined from the equations  
$\varepsilon_{\lambda}(q_{\lambda}(\theta),\theta)=\varepsilon_F$.
Here, $\,k_{so}(\theta)=m^*\Delta(\theta)/\hbar^2$ is the characteristic 
SO momentum, $\varepsilon_F=\hbar^2(k_0^2-2q_{so}^2)/2m^* $ with 
$k_0=\sqrt{2\pi n}$ being the Fermi wave vector of a spin-degenerate 
2DEG with density $n$, and $q_{so}=m^*\sqrt{\alpha^2+\beta^2}/\hbar^2$. 
Because of the SOI, the Fermi line splits into two curves with radii 
given by $q_{\lambda}(\theta)$ which, as the energy surfaces 
$\varepsilon_{\lambda}({\bf k})$, are symmetric with respect to the 
(1,1) and (-1,1) directions in ${\bf k}-$space (Fig.\,1). 
When $\alpha$ or $\beta$ is null, the dispersions are isotropic and 
the Fermi contours are concentric circles.

The 2DEG is excited by a uniform electric field $E$
oscillating at frequency $\omega$ along the $y-$direction. 
The driven electric current is described by the charge current conductivity 
tensor $\sigma_{ij}(\omega)=\delta_{ij}\sigma_D(\omega)+\sigma^s_{ij}(\omega)$,
$\,i,j=x,y$, in which $\sigma_D(\omega)=ine^2/m^*\omega$ is the Drude 
conductivity, and $\sigma^s_{ij}(\omega)$ is the contribution due to
inter-spin-split induced transitions. Within the linear response Kubo 
formalism this SO contribution is 
\begin{equation} \label{Kubo_sigma}
\sigma^s_{ij}(\omega)=\frac{1}{\hbar(\omega+i\eta)}\,\int_0^{\infty}\!dt
\,e^{i(\omega+i\eta)t}\langle[j_i(t),j_j(0)]\rangle \ \ ,
\end{equation}
the symbol $\langle [A(t),B(0)]\rangle=\Sigma_{\lambda}\int^{\prime}d^2 k\, 
f(\epsilon_{\lambda}({\text \bf k}))\langle {\bf k}\lambda|[A(t),B(0)]
|{\bf k}\lambda\rangle$ indicates quantum and thermal 
averaging of the commutator of the operators $A$ and $B$ in the interaction
picture, $f(\varepsilon)$ is the Fermi distribution function, 
and $\eta\to 0^+$. 
The prime on the integral indicates that integration is restricted to 
the region between the Fermi contours, $q_+(\theta)<k<q_-(\theta)$,
for which $\varepsilon_-({\bf k})<\varepsilon_F<\varepsilon_+({\bf k})$,
(Fig. 1). Here, $j_i=ev_i$, with $i=x,y$, is the electric charge current operator,
where $v_i$ is a component of the velocity operator 
${\bf v}({\bf k})=\nabla_{\bf k}H/\hbar=\hbar{\bf k}/m^* + {\bf\hat{x}}
(\beta\sigma_x+\alpha\sigma_y)/\hbar
-{\bf\hat{y}}(\alpha\sigma_x+\beta\sigma_y)/\hbar$.

There is a connection between $\sigma^s_{ij}(\omega)$ and the spin current
response. The spin conductivity describing a $z$-polarized-spin current 
flowing in the $i-$direction as a response to the field $E{\bf\hat{y}}$ is given by

\vspace{-0.15in}
\begin{equation} \label{Kubo_sHall} 
\Sigma^z_{iy}(\omega)=\frac{1}{\hbar(\omega+i\eta)} \int_0^{\infty}\!dt\,
e^{i(\omega+i\eta)t}\langle[{\cal J}^z_i(t),j_y(0)]\rangle \ \ \  ,
\end{equation}
where ${\cal J}^z_i=(\hbar/4)(\sigma_zv_i+v_i\sigma_z)$
is the spin current operator,
with $i$ indicating the transport direction; $\sigma_{x,y,z}$ are the Pauli matrices.

Using the equation of 
motion for $j_y(t)$ to relate $dj_y(t)/dt$ and ${\cal J}^z_x(t)$ 
\cite{Finkelstein} a relation between the above conductivities can be
obtained. In particular, the spin Hall conductivity $\Sigma^z_{xy}$ and the 
diagonal charge conductivity $\sigma^s_{yy}$ are related through the expression
\begin{equation} 
\frac{\Sigma^z_{xy}(\omega)}{(e/8\pi)}=\frac{\hbar\omega}{\varepsilon_R-\varepsilon_D}\,
\frac{i\,\sigma^s_{yy}(\omega)}{(e^2/2\pi\hbar)}\ \ ,
\end{equation}
where $\varepsilon_R=m^*\alpha^2/\hbar^2$ and $\varepsilon_D=m^*\beta^2/\hbar^2$
are the SO characteristic energy scales for the Rashba and Dresselhaus coupling.

\begin{figure}
\includegraphics{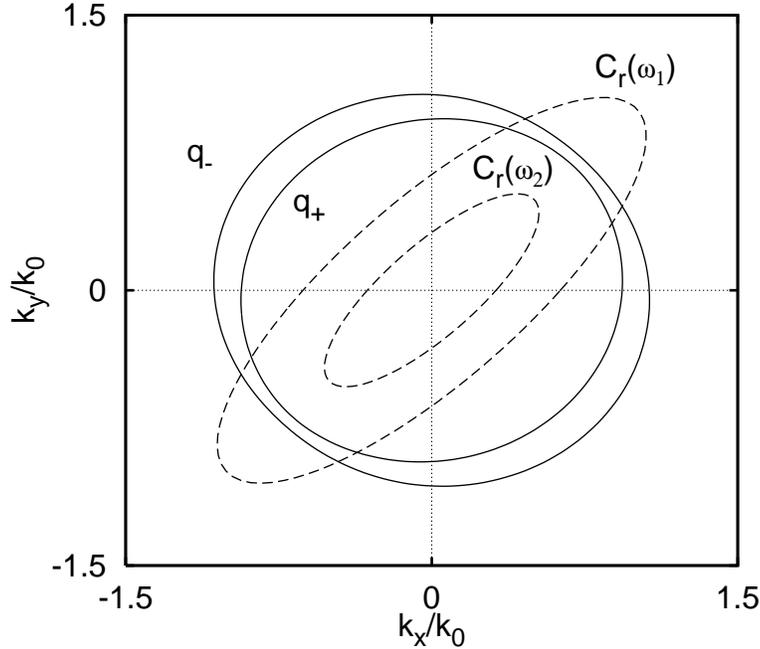}
\caption{Fermi contours $q_{\lambda}(\theta)$ and the 
constant-energy-difference curve $C_r(\omega)$ defined by 
$\varepsilon_+({\bf k})-\varepsilon_-({\bf k})=\hbar\omega$, shown 
for two values of the photon energy $\omega_1>\omega_2$. 
$C_r(\omega)$ is a rotated ellipse 
with semi-axis of lengths $k_a(\omega)=\hbar\omega/2|\alpha-\beta|$ and 
$k_b(\omega)=\hbar\omega/2|\alpha+\beta|$ oriented along the $(1,1)$ and 
$(-1,1)$ directions respectively.
The sample parameters used here are $n=5\times 10^{11}$cm$^{-2}$, 
$\alpha=1.6\times 10^{-9}\,$eV\,cm, $\beta=0.5\alpha$ and $m^*=0.055m$.}
\end{figure}

We evaluate these formulas in the limit of vanishing temperature and 
in the absence of impurity scattering. Given that the optical absorption spectrum
is determined by the imaginary part of the dielectric function
$\epsilon_{ij}(\omega)\propto i\sigma_{ij}(\omega)/\omega$ \cite{Haug_Koch},
in Fig.\,2(a) we show Re$\,\sigma^s_{yy}(\omega)$.
The result is remarkably different from that of the isotropic splitting case 
$\beta = 0, \alpha\neq 0$, for which
$\text{Re}\,\sigma^s_{ij}(\omega)=\delta_{ij}\sigma_R$ for 
$2\alpha k_+\leq\hbar\omega\leq 2\alpha k_-$, where $\sigma_R=e^2/16\hbar$ and 
$k_{\lambda}=q_{\lambda}(\beta=0)$; the absorption band width is thus determined
only by the coupling strength $\alpha$, yielding an spectrum of optical absorption 
essentially featureless.\cite{Xu05,Entin} In contrast,
when both coupling strengths $\alpha$ and 
$\beta$ are present, the spectrum becomes wider and highly asymmetric, and new 
spectral features appear. 
To understand this structure we 
note, from Eq.\,(\ref {Kubo_sigma}), that 
Re$\,\sigma^s_{ij}(\omega)=\int'\!d^2k\,M_{ij}({\bf k}) 
\delta(\varepsilon_+({\bf k})-\varepsilon_-({\bf k})-\hbar\omega)$
with $M_{ij}({\bf k})=(e^2/4\pi\omega) \langle -|v_i({\bf k})|+
\rangle\langle +|v_j({\bf k})|-\rangle$. This expression shows that
the response may be understood in terms of optical transitions 
and energy absorption;
such expression could also be derived from a golden rule calculation.
From the spin splitting of the Fermi line,
it can be seen that the only transitions allowed between spin-split subbands 
$\varepsilon_{\lambda}$ due to photon absorption at energy $\hbar\omega$ 
are those for which $\hbar\Omega_+(\theta)\leq\hbar\omega\leq 
\hbar\Omega_-(\theta)$ with
$\hbar\Omega_+(\theta)=\varepsilon_F-
\varepsilon_-(q_+(\theta),\theta)$ and  $\hbar\Omega_-(\theta)=
\varepsilon_+(q_-(\theta),\theta)-\varepsilon_F$.
That is, for a given $\omega$ only those angular regions in ${\bf k}-$space
satisfying this condition are available for optical transitions (see Fig.\,2(c)).
Thus, the charge conductivity tensor can be recasted as
\begin{eqnarray} \label{Re_sij}
\mbox{Re}\,\sigma^s_{ij}(\omega)&=&\frac{e^2(\alpha^2-\beta^2)^2}
{32\pi\hbar}\! \int\!d\theta\,\frac{\delta_{ij}-(1-\delta_{ij})
\sin\!2\theta}{\Delta^4(\theta)}\,\label{s1} 
\,\Theta[\hbar\omega-\hbar\Omega_+(\theta)]
\Theta[\hbar\Omega_-(\theta)-\hbar\omega], 
\end{eqnarray}
where the unit step functions $\Theta(x)$ impose the above mentioned condition.
This is different to the pure Rashba (or Dresselhaus) case, where 
the whole interval $[0,2\pi]$ contributes to the integral for {\it each}  
allowed photon energy. Interestingly, the non-isotropic spin-splitting 
originated by the simultaneous presence of  both coupling strengths, 
forces the optical excitation to be ${\bf k}-$selective.\cite{RashbaSpinDynamics}
We also notice that for $\alpha^2=\beta^2$ the SO contribution
$\sigma^s_{ij}(\omega)$ vanishes and the charge conductivity tensor becomes
isotropic. This is due to the particular form acquired by the eigenstates and
dispersion relations exactly at that case.\cite{Loss,Egues}

\begin{figure}
\includegraphics{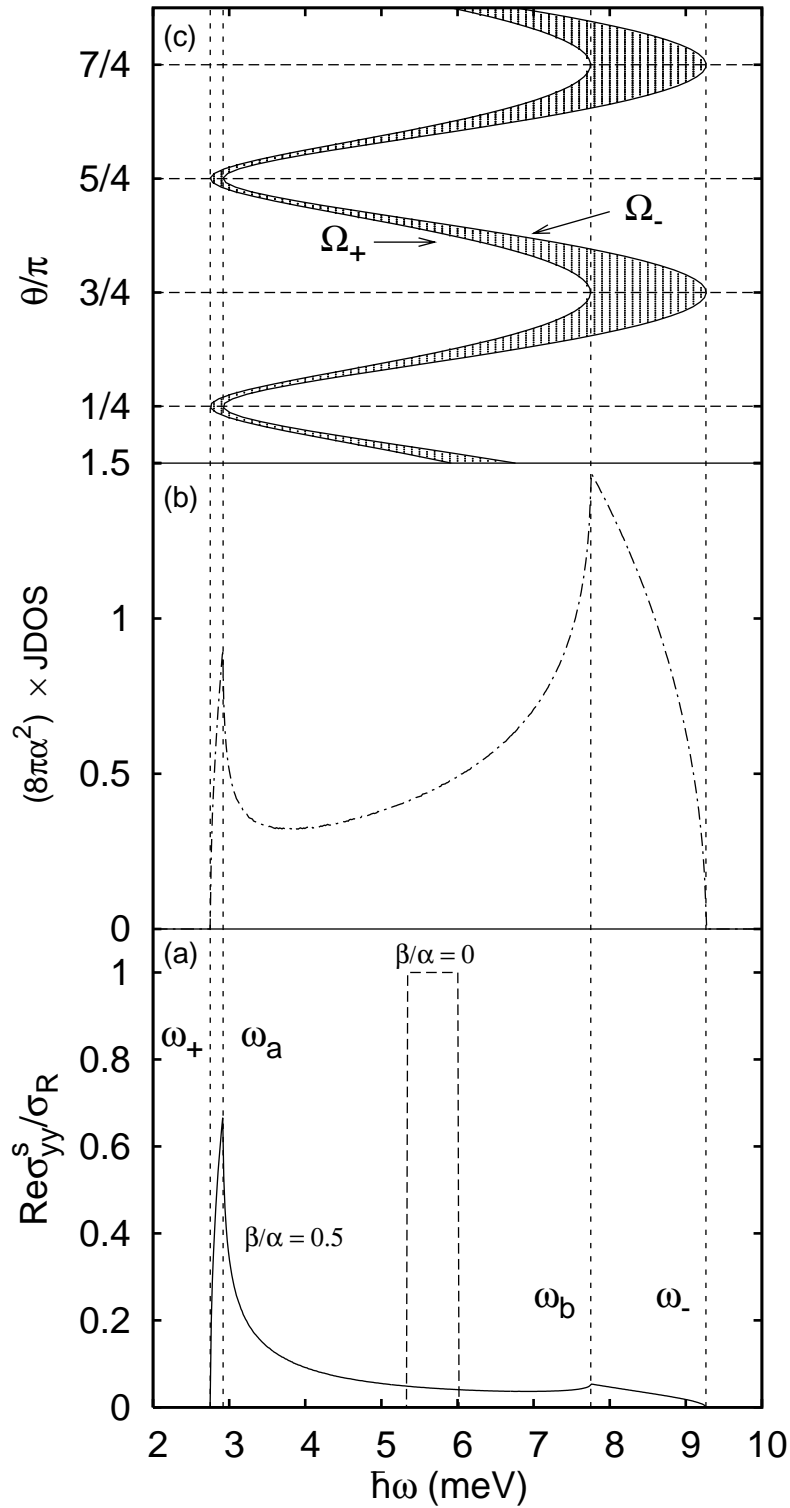}
\caption{(c) Angular region (shaded) in ${\bf k}-$space available
for direct transitions as a function of photon energy. Only the
shaded region contribute to the optical absorption [eq.\,(\ref{Re_sij})]. 
The energy boundaries are given by 
$\hbar\Omega_{\pm}(\theta)=2q_{\pm}(\theta)\Delta(\theta)$.
(b) The joint density of states. (a) SO contribution to the charge
conductivity, Re$\,\sigma^s_{yy}(\omega)$.
For the frequencies $\omega_+=\Omega_+(\pi/4)$, $\,\omega_a=\Omega_-(\pi/4)$,
$\,\omega_b=\Omega_+(3\pi/4)$, $\,\omega_-=\Omega_-(3\pi/4)$, see the text. 
The sample parameters are the same as in Fig.\,1\,.}
\end{figure}

The minimum (maximum) photon energy $\hbar\omega_+\,$($\hbar\omega_-$)
required to induce optical transitions between the initial $\lambda=-1$ 
and the final $\lambda=+1$ subband corresponds to the excitation of an 
electron with wave vector lying on the $q_+\,$($q_-$) Fermi line at 
$\theta_+=\pi/4$ or $5\pi/4$ ($\theta_-=3\pi/4$ or $7\pi/4$), 
giving $\hbar\omega_{\pm}=\hbar\Omega_{\pm}(\theta_{\pm})=
2k_0|\alpha \mp \beta|\mp 2m^*(\alpha \mp \beta)^2/\hbar^2$.
The absorption edges in the spectrum of Fig.2 correspond exactly to 
$\hbar\omega_{\pm}$. The real part of $\sigma^s_{ij}(\omega)$ can also
be written as a line integral to be performed along the arcs of the 
resonant curve $C_r(\omega)$ lying within the region 
enclosed by the Fermi lines $q_{\lambda}(\theta)$ (Fig.\,1). The peaks 
observed in Fig.2(a) correspond to electronic excitations involving 
states with allowed wave vectors on $C_r(\omega)$ such that the velocity 
$|\nabla_{\bf k}(\varepsilon_+-\varepsilon_-)/\hbar|$ takes its minimum
value. The first (second) peak is at a photon energy $\hbar\omega_a$ 
($\hbar\omega_b$) for which the major (minor) semi-axis of the ellipse
$C_r(\omega)$ (Fig.\,1) coincides with the Fermi line $q_-(\theta_+)\,$
($q_+(\theta_-)$),
hence $\hbar\omega_a=\hbar\Omega_-(\theta_+)=2k_0|\alpha-\beta|+
2m^*(\alpha-\beta)^2/\hbar^2$ and $\hbar\omega_b=\hbar\Omega_+(\theta_-)
=2k_0|\alpha+\beta|-2m^*(\alpha+\beta)^2/\hbar^2$. This resembles the
presence of critical points or van Hove singularities which are sources
of structure in the joint density of states (JDOS) and in optical
constants. The JDOS is displayed in Fig.\,2(b). The unequal splitting at the 
Fermi level along the symmetry $(1,1)$ and $(-1,1)$ directions is thus 
responsible for the absorption and high density peaks 
at photon energies $\hbar\omega_a$ and $\hbar\omega_b$ respectively, 
giving meaning to the structure of the spectra. The overall magnitude 
and the asymmetric shape of the spectrum are due to the 
factor $(\alpha^2-\beta^2)^2/\Delta^4(\theta)$ in Eq.(\ref{s1}). 
The results for several values of $\beta/\alpha$ are shown in Fig.\,3.
The value Re$\,\sigma^s_{xy}(\omega)$ (not shown) displays a 
similar spectral behavior as the diagonal component $\sigma^s_{yy}$. 
\begin{figure}
\includegraphics{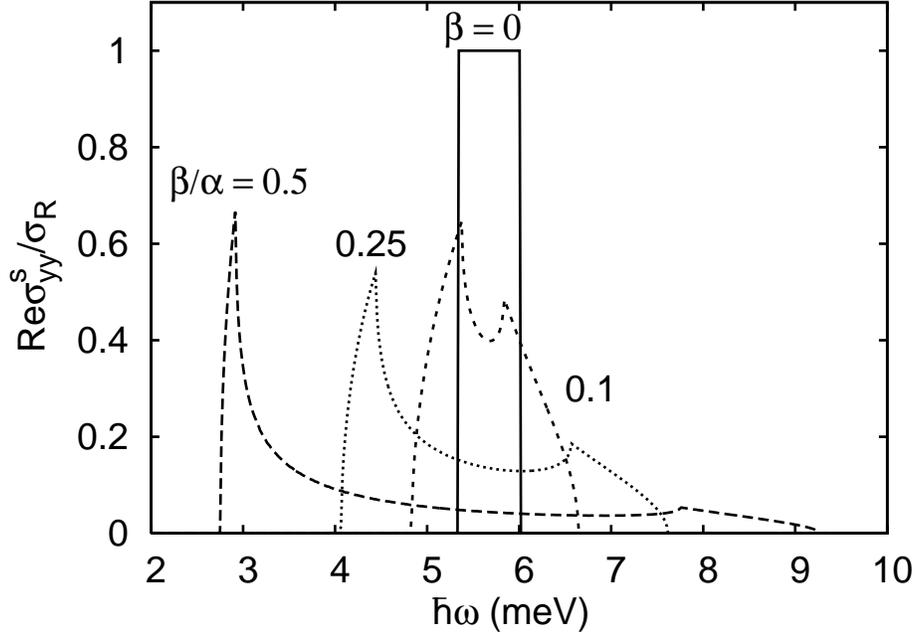}
\caption{Charge conductivity Re$\,\sigma^s_{yy}(\omega)$ for 
several values of the ratio $\beta/\alpha$.
Other parameters are as those used in Fig.\,1\,.}
\end{figure}

The absorption bandwidth $\Delta{\cal E}=\hbar\omega_--\hbar\omega_+$ is 
independent of the frequency and is given by 
\begin{equation}
\Delta{\cal E}(\alpha,\beta;n)=4k_0
[\beta \Theta(\alpha-\beta)+\alpha \Theta(\beta-\alpha)]+ 
4(\varepsilon_R+\varepsilon_D) \ \ ,
\end{equation}
\noindent assuming $(k_{so}(\theta)/k_0)^2\ll 1$. We notice that the two 
first terms can be about an order of magnitude larger than the Rashba 
result, $\Delta{\cal E}_R=4\varepsilon_R$, as the spectra in Fig.\,2 
clearly illustrates for  typical cases. Moreover, $\Delta{\cal E}$ becomes 
explicitly dependent on the electron density $n$ (through the Fermi wave 
vector $k_0$), which is also in contrast to the $\beta=0$ 
case.\cite{Xu05} Thus, as a result of the interplay between the Rashba 
and Dresselhaus interactions, another manipulable parameter ($n$) appears 
to control the optical spin-split response in addition to the tunable coupling 
strength $\alpha$. The expression for $\Delta {\cal E}$ suggests that its 
variation with $\alpha$ can be about an order of magnitude larger if 
$\alpha < \beta$ than for the opposite case. This can be useful to 
determine the sign of $\alpha-\beta$. 

\begin{figure}
\includegraphics{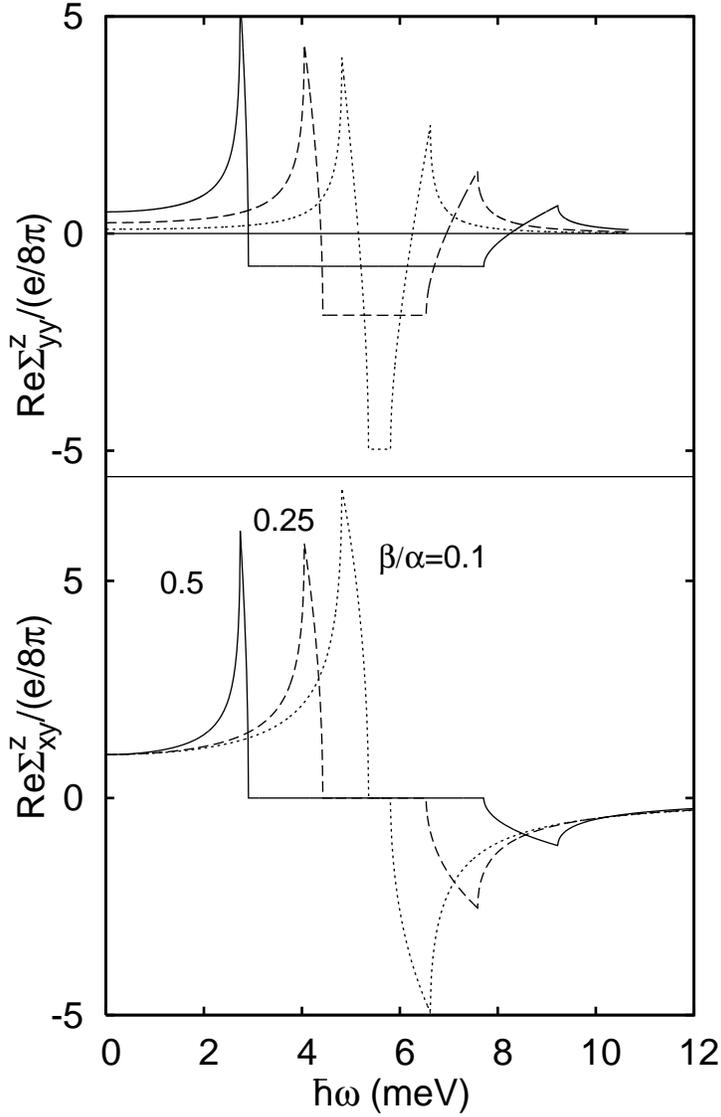}
\caption{Spin Hall conductivity tensor vs. photon energy, 
for $\beta/\alpha =0.5, 0.25, 0.1$ (solid, dashed and dotted respectively). 
The rest of the parameters are the same as in Fig.\,1.}
\end{figure}

Moreover, an expression for the ratio $\beta/\alpha$ can also be derived in 
terms of the relevant frequencies of the absorption spectrum
(assuming $\alpha>\beta>0$),
\begin{equation}
\frac{\beta}{\alpha}=\frac{(\omega_{-}-\omega_{+})+(\omega_{b}-\omega_{a})}
{(\omega_{-}+\omega_{+})+(\omega_{b}+\omega_{a})}
\end{equation} 
which may be useful in optical experiments for extracting 
information of the relative SOI strengths.

As for the spin Hall 
conductivity tensor Re$\Sigma^z_{iy}(\omega)$ we obtain from Eq. 
(\ref{Kubo_sHall})
\begin{equation}\label{Sigma-spin1}
\mbox{Re}\,\Sigma^z_{iy}(\omega)=\Sigma^z_{iy}(0)+
\frac{e}{8\pi}\,\frac{\hbar\omega}{\varepsilon_R-\varepsilon_D}\,I_i(\omega)
\end{equation}

\noindent where explicitly, 
\begin{equation} \label{Sigma-spin}
I_i(\omega)=\frac{(\alpha^2-\beta^2)^2}{8\pi}\,\int_0^{2\pi}\!\!d\theta\,g_i(\theta)\log\!
\left|\frac{[\omega+\Omega_{+}(\theta) ][\omega-\Omega_{-}(\theta) ] } 
{[\omega-\Omega_{+}(\theta) ][\omega+\Omega_{-}(\theta) ]}\right|
\end{equation}
with $g_i(\theta)=[\delta_{ix}\cos^2\!\theta+\delta_{iy} 
\sin\!\theta\cos\!\theta]/\Delta^4(\theta)$.

The static values are 
$\Sigma^z_{xy}(0)=(e/8\pi)$sgn$(\alpha^2-\beta^2)$ and 
$\Sigma^z_{yy}(0)=(e/8\pi)[(\beta/\alpha)\Theta(\alpha^2-\beta^2)-
(\alpha/\beta) \Theta(\beta^2-\alpha^2)]$,
in coincidence with Refs.\,\onlinecite{Sinitsyn} and \onlinecite{SQShen}. 
For $\beta=0$, $I_y(\omega)=0$ and $\Sigma^z_{yy}(\omega)=0$.
As a result, for photon energies $\hbar\omega_a\leq\hbar\omega\leq\hbar
\omega_b$, the tensor Re$\Sigma^z_{iy}(\omega)$ takes a constant value, 
which is zero for the transverse component ($i=x$) and 
$-(e/8\pi)(\alpha^2-\beta^2)/2\alpha\beta$ for the longitudinal component
($i=y$), provided $\alpha\neq\beta\neq 0$. This is a consequence of the 
particular angular symmetry acquired by the integrand in
Eq.\,(\ref{Sigma-spin}) for such frequencies. The spectra in Fig.\,4 
show that the magnitude and the direction  (sign) of
the spin Hall conductivity
$\Sigma^z_{xy}(\omega)$ depend on the frequency and the
SO coupling strengths $\alpha$, $\beta$, and therefore could be manipulated 
via electrical gating and/or by adjusting the light frequency. The latter
offers new possibilities of control of spin currents in electron
systems with competing Rashba and Dresselhaus SOI.

In the above calculations we have not included any type of 
relaxation mechanism.
The parameter $\eta\to 0^+$ means, as usually, that the
perturbation is turned on adiabatically, ensuring a causal response.
In the following, in the line of Ref.\onlinecite{Loss-eta}, we
obtain the static value of the spin conductivity for a finite 
damping parameter $\eta>0$. This parameter accounts phenomenologically for
dissipation effects due to impurity scattering 
and it is considered here as a momentum relaxation rate.

The substitution of $\omega\to i\eta$ \cite{Rashba} in
eqs. (\ref{Sigma-spin1}) and (\ref{Sigma-spin}), leads to

\begin{equation} \label{eta}
\Sigma^z_{iy}(0;\eta)=\Sigma^z_{iy}(0;\eta=0)-\frac{e}{8\pi}\,\frac{\hbar\eta}
{\varepsilon_R-\varepsilon_D}\,F_i(\eta)
\end{equation}
where $\Sigma^z_{iy}(0;\eta=0)$ is the zero frequency value for $\eta\to 0^+$ mentioned
above, and
\begin{equation}
F_i(\eta)=\frac{(\alpha^2-\beta^2)^2}{4\pi}\,\int_0^{2\pi}\!\!d\theta\,g_i(\theta)\,
\arctan\left[\frac{4\varepsilon_{so}(\theta)/\hbar\eta}{1+8\varepsilon_F
\varepsilon_{so}(\theta)/(\hbar\eta)^2}\right]
\end{equation}  
with $\varepsilon_{so}(\theta)=m^*\Delta^2(\theta)/\hbar^2$. It can be shown
that if $m^*(\alpha+\beta)^2/\hbar^2 < \hbar\eta$, eq. (\ref{eta}) 
leads to (for $\alpha\neq\beta\neq 0$)
\begin{eqnarray} 
\Sigma^z_{xy}(0;\eta)&\approx&\frac{e}{\pi}\,\frac{\varepsilon_F}{\hbar\eta}\,
\left(\frac{\varepsilon_R-\varepsilon_D}{\hbar\eta}\right)-
\frac{8e}{\pi}\,\left(\frac{\varepsilon_R+\varepsilon_D}
{\varepsilon_R-\varepsilon_D}\right)\,
\left(\frac{\varepsilon_F}{\hbar\eta}\right)^2\,\left(\frac{\varepsilon_R-\varepsilon_D}
{\hbar\eta}\right)^2 \ \ , \label{eta_1} \\
\Sigma^z_{yy}(0;\eta)&\approx&\frac{8e}{\pi}\,\left(
\frac{\sqrt{\varepsilon_R\varepsilon_D}}{\varepsilon_R-\varepsilon_D}\right)\,
\left(\frac{\varepsilon_F}{\hbar\eta}\right)^2\,\left(\frac{\varepsilon_R-\varepsilon_D}
{\hbar\eta}\right)^2 \ \ . 
\end{eqnarray}
These expressions show that the static limits of the transversal and
longitudinal spin conductivities vanish to different orders in the
parameter $|\varepsilon_R-\varepsilon_D|/\hbar\eta$. For $\beta=0$
equations (\ref{eta}) and (\ref{eta_1}) agree with the results of Ref.\,
\onlinecite{Loss-eta}.

In summary, we have shown that the coexistence of Rashba and Dresselhaus SOI
in 2DEGs induces an anisotropic spin-splitting which gives rise to a 
characteristic frequency dependence  of the charge- and spin
Hall-conductivities of 2DEG systems. Such response provides us 
a `fingerprint signature' of the presence of a competing Rashba and 
Dresselhaus SO mechanism. This also suggests the possibility of the optical 
manipulation of charge and spin (Hall) currents in addition to the control 
obtained through external bias.  Probing the SO coupling strengths through 
optical spectroscopy and/or transport measurements could also be considered.

This work was supported by CONACyT-Mexico 
grants J40521F,J41113F, and by DGAPA-UNAM IN114403-3.


\begin{thebibliography}{00}
\bibitem{Erlingsson-comment}
In Ref.\onlinecite{Erlingsson} $\Sigma^z_{iy}(\omega)$ was obtained to 
lowest order in $q_{so}/k_0\ll 1$, showing resonances at 
$2k_0|\alpha\pm\beta|$. This is equivalent however, to neglecting the terms
$2m^*(\alpha\pm\beta)^2/\hbar^2$ in the resonances $\hbar\omega_a$ and 
$\hbar\omega_b$. They can give non-negligible contributions as our 
calculation illustrates.
\end{thebibliography}
\end{document}